\documentstyle[12pt]{article}
\begin{document}
\oddsidemargin=0.8cm

\def\bib{\bibitem}
\def\be{\begin{equation}}
\def\ee{\end{equation}}
\def\ba{\begin{eqnarray}}
\def\eea{\end{eqnarray}}
\def\df{\stackrel{\rm def}{=} }

\begin{center}
{\Large \bf Gauged Thirring Model in the Heisenberg Picture}
\end{center}
\vspace{1.2cm}
\begin{center}
J. T. Lunardi$\;^1\,$\addtocounter{footnote}{1}\footnote{On leave from Departamento de Matem\'atica e Estat\'{\i}stica, Setor de Ci\^encias Exatas e Naturais, Universidade Estadual de Ponta Grossa, Ponta Grossa, PR, Brazil.}, L. A. Manzoni$\;^2$ and B. M. Pimentel$\;^1$\linebreak[1]
\linebreak[1]
$^1\;$Instituto de F\'{\i}sica Te\'{o}rica\linebreak[1]
Universidade Estadual Paulista\linebreak[1]
Rua Pamplona, 145\linebreak[1]
01405-900 - S\~ao Paulo, S.P. \linebreak[1]
Brazil\linebreak[1]
\linebreak[1]
$^2\;$Instituto de F\'{\i}sica\linebreak[1]
Universidade de S\~ao Paulo\linebreak[1]
Rua do Mat\~ao, Travessa R, 187 \linebreak[1]
Caixa Postal 66318 \linebreak[1]
05315-970 - S\~ao Paulo, S.P. \linebreak[1]
Brazil\linebreak[1]
\linebreak[1]

\end{center}

\vspace{2.2cm}
\begin{abstract}

We consider the (2+1)-dimensional gauged Thirring model in the Heisenberg picture. In this context we evaluate the vacuum polarization tensor as well as the corrected gauge boson propagator and address the issues of generation of mass and dynamics for the gauge boson (in the limits of QED$_3$ and Thirring model as a gauge theory, respectively) due to the radiative corrections.

\end{abstract}

\pagebreak
\section{Introduction}

The Thirring model in (2+1) dimensions has been studied from diverse points of view. In \cite{kky}, for example, the authors have considered this model, which is perturbatively nonrenormalizable, by using the $1/N$ expansion and showed that the model is renormalizable in this context. The model has also been considered to study dynamical generation of a parity-violating mass for fermions
\cite{gom} aiming the comprehension of such a mechanism in four-fermion interactions. This mechanism could be useful for a better understanding of the large mass of top quark.

More recently, the $d$-dimensional ($2\leq d <4$) Thirring model was treated as a gauge theory by making use of hidden local symmetry \cite{ito} and the general formalism of Batalin-Fradkin for a constrained system \cite{kon} (for the abelian model these formalisms are equivalent to the St\"uckelberg procedure \cite{fuj}). The reason for the introduction of a gauge symmetry is that the results for the issue of dynamical mass generation were strongly dependent on the regularization scheme adopted. Then, the existence of a gauge symmetry would be useful in the sense that it restricts the possible regularization schemes. This Thirring model as a gauge theory (TMGT) has revealed a very rich structure and was also used to study dynamical mass generation \cite{ito, kon, sug}.

In the TMGT the vector boson is an auxiliary field at tree level, even though
radiative corrections generate dynamics for it \cite{kon, nos}. So, in
\cite{kon2} Kondo has considered an extension of this model by introducing a
Maxwell term $-\frac{1}{4\gamma}F_{\mu\nu}F^{\mu\nu}$ into the Lagrangian,
obtaining a model which he called gauged Thirring model. This model has the
attractive feature of comprise theories as diverses as QED$_3$ (which is superrenormalizable) and TMGT (nonrenormalizable) in the appropriate limits.

With this in mind, it would be of interest to consider the gauged Thirring model in the light of the Heisenberg picture (HP), once there is no general proof of equivalence between HP and the interaction picture (IP) in quantum field theory, as there is in ordinary quantum mechanics. So, in working with the gauged Thirring model, we can establish the equivalence between these pictures in a class of different theories, including a nonrenormalizable one (TMGT).

Moreover, despite of the great success of the perturbative techniques based on Feynman graphs expansion, we come into difficulties when apply this formalism to nonabelian and nonrenormalizable gauge theories \cite{nak, ana}. This occurs because in the IP formulation of quantum field theories we must split the Lagrangian into free and interacting parts, and none of this parts are invariants under BRST transformations (for a detailed exposition see \cite{nak, ana}). The HP, by its turn, does not suffer of this difficulty because none of such splits is done. Instead we solve the equations of motion for the interacting fields perturbatively, so that these field operators are
given by power series in the coupling constant.

In this work we do not give a complete proof of equivalence of the (2+1)-dimensional gauged Thirring model in the interaction and Heisenberg pictures, instead we just evaluate the one loop vacuum polarization tensor and the corrected gauge boson propagator in HP. In the sequence, we address the important matters of dynamical mass generation (in the case of QED$_3$ limit) and generation of dynamics to the gauge boson (in the limit of TMGT).

By the way, we are also faced another interesting feature of the (2+1)-dimensional theories, namely, the possibility of induction of a Chern-Simons (CS) term \cite{djt} due to radiative corrections. As it is well known, theories involving a CS term have applications in many fields, varying from pure mathematics to condensed matter physics. However, another well stablished fact is that the coefficient of the CS term generated by loops corrections is plagued by a regularization ambiguity \cite{kon, djt}. Here we will adopt the Pauli-Villars regularization \cite{pvi} with a specific choice of the regulators parameters \cite{pto}, so that we obtain a nonvanishing value to the topological mass in accordance with others schemes.

This paper is organized as follows. In Section 2 we will set the notation and review some basic aspects of the gauged Thirring model. Section 3 contains the basic features of perturbation theory in the Heisenberg picture. The vacuum polarization tensor calculations are presented in Section 4. In Section 5 we determine the corrected gauge boson propagator and consider its limits for QED$_3$ and TMGT. Finally, Section 6 is reserved to our concluding remarks.

\section{Gauged Thirring Model}

Let us start recalling the Lagrangian of the original massive Thirring model in (2+1) dimensions:
\begin{equation}
{\cal L} = \overline{\psi}i\gamma^{\mu}\partial_{\mu}\psi  - m\overline{\psi}\psi -\frac{G}{2}(\overline{\psi}\gamma^{\mu}\psi)(\overline{\psi}\gamma_{\mu}\psi)\; ,
\label{l}
\end{equation}

\noindent
with the  algebra for the $\gamma$ matrices in (2+1) dimensions given by
\begin{equation}
\{\gamma^{\mu},\gamma^{\nu}\}=2g^{\mu\nu}, \hspace{0.4cm}
\gamma^{\mu}\gamma^{\nu}=g^{\mu\nu}-i\varepsilon ^{\mu\nu\delta}\gamma_{\delta}\; ,
\label{gcomut}
\end{equation}

\noindent
where $g_{\mu\nu}= {\rm diag} (1, -1, -1)$ and $\varepsilon ^{\mu\nu\delta}$ is the totally antisymmetric Levi-Civita tensor. This algebra can be realized by using the Pauli matrices: $\gamma^0=\sigma_3$, $\gamma^1=i\sigma_1$, $\gamma^2=i\sigma_2$.

In (\ref{l}) $\psi$ is a two component Dirac spinor and the coupling constant $G$ has dimension of ({\it mass})$^{-1}$. For the sake of convenience we will consider the fermion mass positive ($m>0$). This choice does not imply in loss of generality because the fermion mass term breaks parity (besides, it also breaks time reversal). It must be noted that the presence of this term in the Lagrangian induces a Chern-Simons term, which also breaks $P$ and $T$ symmetries, when radiative corrections are taken into account. 

Of course, the model given by Lagrangian (\ref{l}) has no local $U(1)$ invariance. However, after linearizing the interaction by means of an auxiliary vector field, we can use the well known St\"uckelberg procedure \cite{fuj} of introducing a neutral scalar field in such a way that we obtain a gauge invariant theory 
\begin{equation}
{\cal L}^{'} = \overline{\psi}i\gamma^{\mu}(\partial_{\mu}-ieA_{\mu})\psi  - m\overline{\psi}\psi +\frac{M^2}{2}(A_{\mu}-\partial_{\mu}\theta)^2\; ,
\label{ls}
\end{equation}

\noindent
where the scalar mode $\theta$ is the St\"uckelberg field and we have redefined the coupling constant as $G\df \frac{e^2}{M^2}$. Then, it is simple to verify that (\ref{ls}) is invariant under the gauge transformations
\begin{eqnarray}
A_{\mu}&\rightarrow& A^{'}_{\mu}=A_{\mu}+\partial_{\mu}\phi\; ,\nonumber \\ 
\theta &\rightarrow& \theta^{'}=\theta+\phi\; ,\label{gtran} \\ 
\psi &\rightarrow& \psi^{'}=e^{ie\phi}\psi\; .\nonumber 
\end{eqnarray}

The Lagrangian (\ref{ls}), after introducing a gauge fixing and Faddeev-Popov ghost terms, is called the Thirring model as a gauge theory (TMGT) \cite{ito, kon}. An interesting feature of this model is that the gauge boson $A_{\mu}$, which at tree level is an auxiliary field, becomes dynamical due to the radiative corrections \cite{ito, kon, nos}. Nevertheless, it is useful to introduce a gauge invariant kinetic term $ -\frac{1}{4\gamma}F_{\mu\nu}F^{\mu\nu}$ and to analyse the effect of such a term \cite{kon2}. This Maxwell term, besides introducing dynamics for the gauge boson, is also a kind of higher-covariant derivative term and so the theory becomes renormalizable by power counting. Even yet, the resulting theory in not free of ultraviolet divergences and we must use some regularization scheme to perform loop calculations. 

In addition, we can consider the complete BRST symmetric Lagrangian by introducing the gauge fixing and Faddeev-Popov ghost terms so that we obtain, in the $R_{\xi}$ gauge, the so called gauged Thirring model \cite{kon2}
\begin{equation}
{\cal L}_{gTh}={\cal L}_{A,\psi}+{\cal L}_{\theta}+{\cal L}_{gh}\; ,\label{lfim}
\end{equation}

\noindent
with the matter, St\"uckelberg and ghost Lagrangians given, respectively, by
\begin{eqnarray}
{\cal L}_{A,\psi} &=& \overline{\psi}i\gamma^{\mu}D_{\mu}\psi  - m\overline{\psi}\psi -\frac{1}{4\gamma}F_{\mu\nu}F^{\mu\nu}+\frac{M^2}{2}A_{\mu}A^{\mu}-\frac{1}{2\xi}(\partial_{\mu}A^{\mu})^2 \; , \label{lmater}\\ \nonumber \\
{\cal L}_{\theta} &=& \frac{1}{2} (\partial_{\mu}\theta)^2-\frac{\xi M^2}{2} \theta^2\; , \label{lteta}\\ \nonumber \\
{\cal L}_{gh} &=& i\left[ (\partial_{\mu}\overline{c})(\partial^{\mu}c)- \xi M^2 \overline{c}c\right]\; . \label{lghost}
\end{eqnarray}

\noindent
where the covariant derivative is $D_{\mu}=\partial_{\mu}-ieA_{\mu}$.

From (\ref{lmater})-(\ref{lghost}) we see that just the fermion fields and the gauge boson ones are interacting. Thus, we will concentrate on the matter Lagrangian (\ref{lmater}). In addition, it should be stressed that, in spite of the mass term for the vector field in (\ref{lmater}), in this case $A_{\mu}$ is a true gauge field due to the introduction of St\"uckelberg's field. In fact, the Lagrangian (\ref{lfim}) would be obtained more directly by starting from the Lagrangian of a massive vector field (the Proca field) and applying the St\"uckelberg formalism. However, the procedure outlined above is useful to illustrate the connection between this model and the Thirring one. 

It is worth to consider some limits of the gauged Thirring model. Firstly, in the $\gamma \rightarrow \infty$ the boson kinetic term vanishes and the TMGT, where the gauge boson is an auxiliary field, is recovered. If, in addition, we take the unitary gauge $\xi\rightarrow \infty$, we recover the original Thirring model. Another limit of interest is that in which $\gamma\rightarrow 1$ and the
gauge boson mass goes to zero $M\rightarrow 0$ (which, with $e$ maintained fixed, corresponds to the limit of strong Thirring coupling constant $G\rightarrow \infty$), when the matter Lagrangian ${\cal L}_{A,\psi}$ goes to the QED$_3$ Lagrangian.

\section{Perturbation Theory in the Heisenberg Picture}

In the Heisenberg picture the Lagrangian (\ref{lfim}) is not split into its free and interacting parts, in contradistinction to the interaction picture approach. Instead, we derive via variational principle, the coupled equations of motion of the interacting fields $\psi$ and $A_{\mu}$. In this section we shall be concerned with perturbative solutions of these equations.

In order to consider the vacuum polarization, which will be done with details in the next section, it is convenient to add to the Lagrangian a term
$e\overline{\psi}\gamma^{\mu}\psi A^{\rm ext}_{\mu}$, corresponding to an applied external field. So, we obtain the following equations of motion
\ba
&&(i\partial\!\!\!\slash -m)\psi (x)=-e\{ A\!\!\!\slash (x) + A\!\!\!\slash^{\rm ext}(x)\} \psi (x)\; ; \label{eqpsi} \\ \nonumber \\
&&\left(\Box +\gamma M^2\right) A^{\mu}(x) +\left( \frac{\gamma}{\xi}-1\right) \partial^{\mu}\{ \partial_{\nu}A^{\nu}(x)\} = \gamma j^{\mu}(x)\; ,\label{eqa}
\eea

\noindent
with 
\be
j^{\mu}(x)= -e\overline{\psi}(x)\gamma^{\mu}\psi (x)\; .
\ee

These equations can be put into an integral form by the standard method of Green's functions: 
\ba
\psi (x)&=&\psi^{({\rm in})}(x)+\int d^3y S_{\rm R}(x-y)e(A\!\!\!\slash (y) +
A\!\!\!\slash^{\rm ext}(y))\psi (y)\; ; \label{eqintpsi} \\ \nonumber \\
A_{\mu}(x)&=&A_{\mu}^{({\rm in})}(x)-\int d^3y D_{\rm R}(x-y)\gamma j_{\mu}(y)\; , \label{eqinta}
\eea

\noindent
where $\psi^{({\rm in})}(x)$ and $ A_{\mu}^{({\rm in})}(x)$ are solutions of the free equations of motion and satisfy the well known free field (anti)commutation relations. $S_{\rm R}(x)$ and $D_{\rm R}(x)$ are the retarded Green's functions. For later use, we explicit the form of $S_{\rm R}(x)$ and also the advanced function $S_{\rm A}(x)$:
\ba
S_{\rm R}(x)\!\!&=&\!\!\frac{1}{(2\pi )^3}\int d^3 p (p\!\!\!\slash +m) \left\{ {\rm PV}\frac{1}{m^2-p^2}+ i\pi {\rm sgn}(p^0) \delta(m^2-p^2)\right\} e^{-ip\cdot x};  \label{sr} \\ \nonumber  \\
S_{\rm A}(x)\!\!&=&\!\!\frac{1}{(2\pi )^3}\int d^3 p (p\!\!\!\slash +m) \left\{ {\rm PV}\frac{1}{m^2-p^2}- i\pi {\rm sgn}(p^0) \delta(m^2-p^2)\right\} e^{-ip\cdot x}.  \label{sa}
\eea

The equations (\ref{eqintpsi}) and (\ref{eqinta}) can be solved by iteration, so that we can write the fields $\psi (x)$ and $A_{\mu}(x)$ as expansions in powers of the coupling constant $e$ (and the external field $A_{\mu}^{\rm ext}(x)$). Thus
\ba
\psi (x)&=&\psi^{({\rm in})}(x)+e\int d^3y S_{\rm R}(x-y)(A\!\!\!\slash^{\rm (in)}(y) +
A\!\!\!\slash^{\rm ext}(y))\psi^{(\rm in)} (y) + \cdots\; ; \label{pexp} \\ \nonumber \\
A_{\mu}(x)&=&A_{\mu}^{({\rm in})}(x)+e\gamma \int d^3y D_{\rm R}(x-y)\overline{\psi}^{\rm (in)}(y)\gamma_{\mu}\psi^{\rm (in)} (y) + \cdots \; .\label{aexp}
\eea

\noindent
Taking the adjoint of (\ref{pexp}) we get
\be
\overline{\psi} (x)=\overline{\psi}^{({\rm in})}(x)+e\int d^3y \overline{\psi}^{(\rm
in)}(y)(A\!\!\!\slash^{\rm (in)}(y) +
A\!\!\!\slash^{\rm ext}(y))S_{\rm A}(y-x) + \cdots\; . \label{pexpa} 
\ee

In order to eliminate the zero-point charge we renormalize the current by redefining it as
\be
j^{\mu}(x)=-\frac{e}{2}[\overline{\psi}(x),\gamma^{\mu}\psi (x)]\; ,
\ee

\noindent
where $[\overline{\psi},\gamma^{\mu}\psi
]=\gamma^{\mu}_{\alpha\beta}[\overline{\psi}_{\alpha},\psi_{\beta}]$ (this could be done from the very beginning by the appropriate symmetrization of the operator Lagrangian \cite{kal}). Then, substituting the expansion (\ref{pexp}) into this expression we have
\ba
j^{\mu}(x)&=&-\frac{e}{2}[\overline{\psi}^{(\rm in )}(x),\gamma^{\mu}\psi^{(\rm in )} (x)]-\frac{e^2}{2}\int d^3y \left\{ [\overline{\psi}^{(\rm in )}(x),\gamma^{\mu}S_{\rm R}(x-y)\gamma^{\nu}\psi^{(\rm in )} (y)] \right. \nonumber \\ \nonumber \\
&+&\left. [\overline{\psi}^{(\rm in )}(y)\gamma^{\nu}S_{\rm A}(y-x),\gamma^{\mu}\psi^{(\rm in )} (x)] \right\} \left\{A_{\nu}^{(\rm in)}(y)+A_{\nu}^{\rm ext}(y)\right\} + \cdots\; .
\eea

\noindent
Taking the vacuum expectation value of this current we obtain 
\be
\langle 0| j^{\mu}(x)|0\rangle =\int d^3y \; \Pi^{\mu\nu}(x-y)A_{\nu}^{\rm ext}(y)\; ,
\ee

\noindent
with
\ba
\Pi^{\mu\nu}(x-y)&=&-\frac{e^2}{2}{\rm Tr} \left\{  \gamma^{\mu}S_{\rm R}(x-y)\gamma^{\nu}S^{(1)}(y-x) \right. \nonumber \\ \label{picoor}\\
&+&\left.  \gamma^{\nu}S_{\rm A}(y-x)\gamma^{\mu}S^{(1)}(x-y)\right\}\; , \nonumber 
\eea

\noindent
where 
\ba
S^{(1)}_{\alpha\beta}(x-y)&=& \langle 0 |\; [ \;\overline{\psi}_{\beta}^{(\rm in)}(y), \psi_{\alpha}^{(\rm in)}(x) ]\; |0 \rangle \nonumber \\ \label{s1} \\
&=& \frac{1}{(2\pi )^2}\int d^3p\; e^{-ip\cdot (x-y)}(p\!\!\!\slash +m)_{\alpha\beta} \delta(p^2-m^2)\; . \nonumber
\eea 

As we see, a nonvanishing vacuum expectation value of the current is induced only if an external field is applied. Then, the vacuum behaves like a polarizable medium and the kernel $\Pi^{\mu\nu}(x-y)$ is the vacuum polarization tensor, considered in that follows.

\section{The Vacuum Polarization Tensor}

Now we shall be concerned with the calculation of $\Pi^{\mu\nu}$. After substituting (\ref{sr}), (\ref{sa}) and (\ref{s1}) into (\ref{picoor}) we obtain its Fourier transform 
\ba
\Pi^{\mu\nu}(k) &=& -\frac{e^2}{2(2\pi )^2}\int d^3p_1 d^3p_2\; \delta (k-p_1+p_2) \nonumber \\ \label{hpol} \\
&\times& {\rm Tr} \{ \gamma^{\mu}({p\!\!\!\slash}_1 +m)\gamma^{\nu}({p\!\!\!\slash}_2 +m)\} \left[ \Pi^+ (p_1,p_2)+ \Pi^- (p_1,p_2)\right] \; , \nonumber
\eea

\noindent
where
\ba
\Pi^+(p_1,p_2)&\df& \delta (m^2-p_2^2)\left\{ {\rm PV}\frac{1}{m^2-p_1^2}+ i\pi {\rm sgn}(p_1^0) \delta(m^2-p_1^2)\right\} \; ; \nonumber \\  \\
\Pi^-(p_1,p_2)&\df& \delta (m^2-p_1^2)\left\{ {\rm PV}\frac{1}{m^2-p_2^2}- i\pi {\rm sgn}(p_2^0) \delta(m^2-p_2^2)\right\} \; . \nonumber
\eea

\noindent
Performing the trace and $p_2$ integration in (\ref{hpol}) we obtain
\ba
\Pi^{\mu\nu}(k) &=& -\frac{e^2}{(2\pi )^2}\int d^3p_1 \left\{ p_1^{\mu}(p_1-k)^{\nu}+ p_1^{\nu}(p_1-k)^{\mu} +g^{\mu\nu}(m^2-p_1^2+p_1\cdot k) \right. \nonumber \\ \nonumber\\
&+& \left. im\varepsilon^{\mu\nu\delta}k_{\delta} \right\} \left[ \Pi^+ (p_1,p_1-k)+ \Pi^- (p_1,p_1-k)\right]\; . \label{poltr}
\eea

Now, by using the gauge invariance of the model and Lorentz covariance, we expect that the vacuum polarization tensor can be written in the form
\begin{equation}
\Pi^{\mu\nu}(k)=\left(g^{\mu\nu}-\frac{k^{\mu}k^{\nu}}{k^2}\right)\Pi^{(1)}(k^2)+im\varepsilon^{\mu\nu\delta}k_{\delta}\Pi^{(2)}(k^2)\; ,
\label{polageral}
\end{equation}

\noindent
with the form factors given by

\ba
\Pi^{(1)}(k^2) &=& \frac{1}{2}\Pi^{\mu}_{\; \mu}(k) \; ;\label{pi1} \\ \nonumber \\
\Pi^{(2)}(k^2) &=& -\frac{i}{2m}\varepsilon_{\mu\nu\delta}\frac{k^{\delta}}{k^2}\Pi^{\mu\nu}(k)\; . \label{pi2}
\eea

Thus, the next step would be turned equation (\ref{poltr}) into (\ref{pi1}) and (\ref{pi2}). However, we must be careful because the integral in (\ref{poltr}) is a divergent one. So, it needs to be regularized. Here we will adopt Pauli-Villars (PV) regularization, that preserves gauge invariance and it is simple to implement in the context of Heisenberg picture.

Then the regularized vacuum polarization tensor is

\be
\Pi^{\mu\nu}_{\rm reg} (k) =\sum_{i=0}^n C_i\; \Pi^{\mu\nu}(k,m_i)\; ,
\ee

\noindent
where $\Pi^{\mu\nu}(k,m_i)$ is the expression (\ref{poltr}) with $m$ replaced by $m_i$ and $n$ is the number of auxiliary spinors fields sufficient to remove the divergences. For this purpose the coefficients $C_i$'s must satisfy the consistency conditions:
\ba
&&\sum_{i=0}^n C_i =0\; ; \nonumber \\ \label{ci} \\
&&\sum_{i=0}^n C_i m_i =0 \; ,\nonumber
\eea

\noindent
with $C_0=1$ and $m_0=m$. By analysing (\ref{poltr}) we see that to remove the divergences of the vacuum polarization tensor we need just two regulator fields, so that henceforward we will take $C_j=0$ for all $j> 2$.

Then, considering $\Pi^{\mu\nu}_{\rm reg} $, it is possible to consider separately the form factors according to (\ref{pi1}) and (\ref{pi2}) (which are the symmetric and antisymmertic parts of vacuum polarization, respectively). Firstly we shall consider the antisymmetric part because this is related with the topological mass and it is here that the regularization controversy emerges. For pratical pourposes we will split $\Pi^{(2)}_{\rm reg}(k^2) $ in its real and imaginary parts
\be
\Pi^{(2)}_{\rm reg}(k^2)= \Re {\rm e}\;  \Pi^{(2)}_{\rm reg} (k^2)  + i\; \Im {\rm m}\; \Pi^{(2)}_{\rm reg} (k^2)\; ,
\ee

\noindent
where
\ba
\Re {\rm e}\;  \Pi^{(2)}_{\rm reg} (k^2)&=& -\frac{e^2}{(2\pi )^2m}\sum_i C_im_i \int d^3 p \left\{ \delta (m_i^2-p^2)\; {\rm PV}\frac{1}{m_i^2-(p-k)^2} \right. \nonumber \\ \label{real} \\
&+& \left. \delta \left[ m_i^2 -(p-k)^2\right] {\rm PV} \frac{1}{m_i^2-p^2} \right\}\; ; \nonumber \\ \nonumber \\
\Im {\rm m}\; \Pi^{(2)}_{\rm reg} (k^2) &=& -\frac{e^2}{4\pi m}\sum_i C_i m_i\int d^3 p \; \delta (m_i^2-p^2)\delta \left[ m_i^2 -(p-k)^2\right] \nonumber \\ \label{imag} \\
&\times& \left\{ {\rm sgn}(p^0)-{\rm sgn}(p^0-k^0)\right\}\; . \nonumber
\eea

For na\"{\i}ve power counting these integrals are finite, so we would expect that the result to the form factor $\Pi^{(2)}$ would be independent of the regularization scheme. Nevertheless, it is well known that there is a regularization ambiguity related to this term \cite{kon, djt}. In particular, it is generally accepted that PV regularization gives no mass correction, in contradistinction to analytic \cite{pst} and dimensional \cite{djt} ones. This happens because in the PV scheme the effect of regulator fermionic masses remains finite even after we take the limit of these regulators going to infinity (the result depend on the masses signals). However, it is possible to make a specific choice of the regulators such that the finite integrals remains unaffected by the PV regularization \cite{pto}.

Let us now see how this works. The real part of $\Pi^{(2)}_{\rm reg}$ can be written, by using well known representations for the principal value and delta function \cite{kal}, as
\be
\Re {\rm e}\;  \Pi^{(2)}_{\rm reg} (k^2) = \frac{ie^2}{m(2\pi )^3}\sum_i C_i m_i\int_0^1d\alpha \int_{-\infty}^{\infty}dx\; x \int d^3p\; e^{i x (M^2_i-p^2)}\; ,
\label{reint}
\ee

\noindent
where we have defined
\be
M^2_i\df m^2_i-\alpha(1-\alpha)k^2\; .
\ee

The $x$ integral can be performed by mean of the usual Fourier transform of the $x^{\lambda}_+$ \cite{gel} and leads to
\ba
\Re {\rm e}\;  \Pi^{(2)}_{\rm reg} (k^2) &=& - \frac{ie^2}{m(2\pi )^3}\sum_i C_i m_i\int_0^1 d\alpha\int d^3p \left\{ \frac{1}{[M^2_i-p^2+i\epsilon]^2} \right. \nonumber \\ \\
&-& \left. \frac{1}{[M^2_i-p^2-i\epsilon]^2} \right\} \; , \nonumber
\eea

\noindent
these integrals can be solved by standard techniques and results
\begin{equation}
\Re {\rm e}\; \Pi^{(2)}_{\rm reg}(k^2) = \frac{e^2}{4\pi m\sqrt{k^2}} \sum_i C_i m_i\ln \left| \frac{1-\sqrt{\frac{k^2}{4m^2_i}}}{1+\sqrt{\frac{k^2}{4m^2_i}}}\right|\; .
\end{equation}

\noindent
Now, taking the masses of the regulator fields going to infinity we get
\ba
\Re {\rm e}\; \Pi^{(2)}_{\rm reg}(k^2) &=& \frac{e^2}{4\pi \sqrt{k^2}} \ln \left| \frac{1-\sqrt{\frac{k^2}{4m^2}}}{1+\sqrt{\frac{k^2}{4m^2}}}\right| \nonumber \\ \label{pi2-reg-res} \\
&-& \frac{e^2}{4\pi m}\left[ C_1{\rm sgn}(m_1)+C_2{\rm sgn}(m_2)\right] \; ,\nonumber
\eea

\noindent
and it is easy to show that with the choice $C_1=C_2=-\frac{1}{2}$ we have $m_2=-m_1$ \cite{pto}, so that the term in the square brackets in (\ref{pi2-reg-res}) cancels and we obtain
\be
\Re {\rm e}\; \Pi^{(2)}_{\rm reg}(k^2)\rightarrow \Re {\rm e}\; \Pi^{(2)}(k^2) = \frac{e^2}{4\pi \sqrt{k^2}} \ln \left| \frac{1-\sqrt{\frac{k^2}{4m^2}}}{1+\sqrt{\frac{k^2}{4m^2}}}\right|\; .
\label{pi2-res}
\ee

The calculation of the imaginary part, equation (\ref{imag}), is straightforward and one obtains 
\begin{equation}
\Im {\rm m}\; \Pi^{(2)}_{\rm reg}(k^2) \rightarrow \Im {\rm m}\; \Pi^{(2)}(k^2) = -\frac{e^2}{4\sqrt{k^2}}\;{\rm sgn}(k_0) \Theta (k^2-4m^2)\; .
\label{pi2-res-im}
\end{equation}

\noindent
From (\ref{pi2-res}) and (\ref{pi2-res-im}) we obtain that the coefficient of antisymmetric part at zero momentum is given by

\be
m\Pi^{(2)}(0) = -\frac{e^2}{4\pi} \; ,
\label{pi2em0}
\ee

\noindent
which by the Coleman-Hill theorem \cite{chi} is the exact value for the topological mass, since the higher order contributions to vacuum polarization tensor do not contribute to this term. Then, we see that by an appropriate choice of regulator parameters the PV regularization results in the same value for $\Pi^{(2)}(0)$ that analytic \cite{pst} and dimensional \cite{djt} ones.

Turning now to the symmetric form factor, equation (\ref{pi1}), we can also split it in its real and imaginary parts as 
\be
\Pi^{(1)}_{\rm reg}(k^2)= \Re {\rm e}\;  \Pi^{(1)}_{\rm reg} (k^2)  + i\; \Im {\rm m}\; \Pi^{(1)}_{\rm reg} (k^2)\; ,
\ee

\noindent
where
\ba
\Re {\rm e}\;  \Pi^{(1)}_{\rm reg} (k^2)&=& -\frac{e^2}{2(2\pi )^2}\sum_i C_i \int d^3 p \left\{ \delta (m_i^2-p^2)\; {\rm PV}\frac{1}{m_i^2-(p-k)^2} \right. \nonumber \\ \label{real1} \\
&+& \left. \delta \left[ m_i^2 -(p-k)^2\right] {\rm PV} \frac{1}{m_i^2-p^2} \right\} [3m^2_i-p\cdot (p-k)]\; ; \nonumber \\ \nonumber \\
\Im {\rm m}\; \Pi^{(1)}_{\rm reg} (k^2) &=& -\frac{e^2}{8\pi}\sum_i C_i \int d^3 p \; \delta (m_i^2-p^2)\delta \left[ m_i^2 -(p-k)^2\right] \nonumber \\ \label{imag1} \\
&\times& \left\{ {\rm sgn}(p^0)-{\rm sgn}(p^0-k^0)\right\}[3m^2_i-p\cdot (p-k)]\; . \nonumber
\eea

\noindent
The calculation of these expressions follows along the same lines outlined above for the antisymmetric part (with minor modifications), so we just write down the result for the complete $\Pi_{\rm reg}^{(1)}(k^2)$:
\begin{eqnarray}
\Pi_{\rm reg}^{(1)}(k^2)\rightarrow \Pi^{(1)}(k^2)
&=& \frac{e^2}{16\pi}k^2 \left[\frac{4m}{k^2} +\frac{1}{\sqrt{k^2}} \left(1+\frac{4m^2}{k^2}\right) \right. \nonumber \\ \nonumber \\
&\times& \left. \left( \ln \left| \frac{1-\sqrt{\frac{k^2}{4m^2}}}{1+\sqrt{\frac{k^2}{4m^2}}}\right|- i\pi{\rm sgn}(k_0) \Theta (k^2-4m^2)\right) \right]\; ,
\label{pi1-res}
\end{eqnarray}

\noindent
from which we obtain
\be
\Pi^{(1)}(0)=0\; .
\label{pi1em0}
\ee

\section{Corrected Gauge Boson Propagator}

Now we can obtain the gauge boson propagator corrected by vacuum polarization insertions to one loop, which enable us to account the subjects of dynamical mass generation and generation of dynamics in the QED$_3$ and TMGT limits, respectively.

The corrected gauge boson propagator is given by 
\begin{equation}
{\cal D}^{-1}_{\mu\nu}= (D_F^{\mu\nu})^{-1}+\Pi^{\mu\nu}.
\label{propcompleto}
\end{equation}

\noindent
where $D_F^{\mu\nu}$ is the free propagator
\be
D^{\mu\nu}_F(k)= -\frac{1}{\frac{1}{\gamma} k^2-M^2}\left( g_{\mu\nu} -(1-\frac{\xi}{\gamma} )\frac{k_{\mu}k_{\nu}}{k^2 - \xi M^2} \right)\; ,
\label{free}
\ee

To perform the inversions above it is convenient to introduce the complete set of orthonormal projectors
\begin{eqnarray}
P^{\mu\nu}_{(1)}&=& \frac{1}{2}\left( g^{\mu\nu}-\frac{k^{\mu}k^{\nu}}{k^2}+i\varepsilon^{\mu\nu\delta}\frac{k_{\delta}}{\sqrt{k^2}}\right), \nonumber \\ \nonumber \\
P^{\mu\nu}_{(2)}&=& \frac{1}{2}\left( g^{\mu\nu}-\frac{k^{\mu}k^{\nu}}{k^2}-i\varepsilon^{\mu\nu\delta}\frac{k_{\delta}}{\sqrt{k^2}}\right),  \\ \nonumber \\
P^{\mu\nu}_{(3)}&=&\frac{k^{\mu}k^{\nu}}{k^2}\; ,\nonumber
\label{projetores}
\end{eqnarray}

\noindent
in terms of which the vacuum polarization tensor is given by
\begin{equation}
\Pi^{\mu\nu}(k)=( P^{\mu\nu}_{(1)}+ P^{\mu\nu}_{(2)})\Pi^{(1)}(k^2) +m\sqrt{k^2}( P^{\mu\nu}_{(1)}- P^{\mu\nu}_{(2)})\Pi^{(2)}(k^2)\; .
\end{equation}

To make the inversion of the propagator we just write it using these projectors and invert the respective coefficients. So, we have
\begin{equation}
(D_F^{\mu\nu})^{-1}= -(\frac{1}{\gamma} k^2-M^2)\left( P^{\mu\nu}_{(1)} + P^{\mu\nu}_{(2)}\right) -\frac{k^2-\xi M^2}{\xi} P^{\mu\nu}_{(3)} \; .
\end{equation}

Substituting these expressions into (\ref{propcompleto}) and inverting again, we finally obtain the corrected propagator
\begin{eqnarray}
{\cal D}^{\mu\nu}(k)&=&\frac{1}{k^2-\tilde{\Pi}(k^2)}\left[ \left( g^{\mu\nu}-\frac{k^{\mu}k^{\nu}}{k^2}\right)\frac{[\frac{1}{\gamma} k^2 - M^2 -\Pi^{(1)}(k^2)]}{[m\Pi^{(2)}(k^2)]^2}+    i\varepsilon^{\mu\nu\delta}\frac{k_{\delta}}{m\Pi^{(2)}(k^2)}\right] \nonumber \\ \nonumber \\
&-&\xi\frac{k^{\mu}k^{\nu}}{k^2(k^2-\xi M^2)}\; ,
\label{propcomp}
\end{eqnarray}

\noindent
where we have defined
\begin{equation}
\tilde{\Pi}(k^2)\df \frac{[\frac{1}{\gamma} k^2 -M^2-\Pi^{(1)}(k^2)]^2}{[m\Pi^{(2)}(k^2)]^2}\; .
\label{pole}
\end{equation}

From equation (\ref{propcomp}) we see that the corrected gauge boson propagator acquires an antisymmetric parity-violating part correspondent to the induction of a CS term. It must be observed that this expression has a well defined limit as $m \rightarrow 0$,  but in this case we do not have a CS term.

Let us now to analyse some limits of (\ref{propcomp}). First, let us make $\gamma\rightarrow\infty$, when the Lagrangian (\ref{lfim}) goes to the TMGT Lagrangian. This limit is easily performed in (\ref{propcomp})-(\ref{pole}) and we can observe that a pole is generated in the corrected propagator (note that the free propagator, equation (\ref{free}), does not have a pole in the same limit). This indicates that the gauge boson of TMGT becomes dynamical due to radiative corrections \cite{ito, kon, nos}.

Another limit of interest is given by taking $\gamma\rightarrow 1$ and the gauge boson mass vanishing $M\rightarrow 0$ (i.e., the strong Thirring coupling constant limit $G\rightarrow \infty$). In this limit the matter sector of the gauged Thirring model goes to QED$_3$. After taking this limit (\ref{propcomp}) can be written in the form 
\begin{eqnarray}
\left.{\cal D}^{\mu\nu}(k)\right|_{\stackrel{M^2\rightarrow \;0}{\gamma \rightarrow 1}}&=&-\frac{1}{k^2-\Pi(k^2)}\left[ \left( g^{\mu\nu}-\frac{k^{\mu}k^{\nu}}{k^2}\right)+    im\varepsilon^{\mu\nu\delta}\frac{k_{\delta}}{k^2}\frac{\Pi^{(2)}(k^2)}{1-\frac{\Pi^{(1)}(k^2)}{k^2}}\right] \nonumber \\ \nonumber \\
&-&\frac{\xi}{k^2}\frac{k^{\mu}k^{\nu}}{k^2}\; ,
\label{qed3}
\end{eqnarray}

\noindent
where
\begin{equation}
\Pi(k^2)=\Pi^{(1)}(k^2)+ \frac{ [m\Pi^{(2)}(k^2)]^2}{1-\frac{\Pi^{(1)}(k^2)}{k^2}} \; .
\label{p}
\end{equation}

The propagator (\ref{qed3}) has the same form as that obtained in ref. \cite{djt}. Then, the behaviour of $\Pi (0)$ dictates whether a mass for the gauge boson is generated or not. On the other hand, as $\Pi^{(1)}(0)=0$ (equation (\ref{pi1em0})), we see by (\ref{p}) that this question depends exclusively on the behaviour of $\Pi^{(2)}(0)$. The usual treatments using Pauli-Villars regularization give $\Pi^{(2)}(0)=0$, so that the gauge boson remains massless. However, with the choice adopted for the $C_i$'s we get $\Pi^{(2)}(0)\neq 0$ (see equation (\ref{pi2em0})) and so the gauge boson of QED$_3$ acquires a mass, in accordance with the results obtained by using dimensional \cite{djt} or analytic \cite{pst} regularization or the Epstein and Glaser causal method \cite{swp}.

\section{Concluding Remarks}

In the framework of the Heisenberg picture we have calculated the vacuum polarization tensor in the (2+1)-dimensional gauged Thirring model by using the Pauli-Villars regularization. By an appropriate choice of the regulators \cite{pto} we were able to obtain a nonvanishing value for the coefficient of the induced Chern-Simons term (topological mass), which is in accordance with the result obtained by others regularization schemes \cite{djt, pst}. We must stress that the choice of the regulators which we have made does not modify the finite parts of the amplitude, a highly desirable feature in a regularization. In this sense that choice is the most natural one \cite{pto}.

By\hfill considering\hfill the\hfill corrected\hfill gauge\hfill boson\hfill propagator\hfill and\hfill taking\hfill the\hfill limit\hfill\\
$\gamma\rightarrow\infty$, when the gauged Thirring model goes to the TMGT, we have reproduced the results obtained working with the TMGT Lagrangian from the very beginning \cite{nos}. Thus, we see that the gauge boson of this model becomes dynamical by radiative corrections. In the limit of strong Thirring coupling constant $M\rightarrow 0$ (with $\gamma =1$) the gauged Thirring model goes to the QED$_3$. Then, from the corrected propagator we see that the photon becomes massive due to the nonvanishing value of $\Pi^{(2)}(0)$ \cite{djt}.

In special, this example is very interesting since the gauged Thirring model comprises theories as distincts as QED$_3$ (superrenormalizable) and the Thirring model as a gauge theory (nonrenormalizable) as limits. Work considering the complete one loop treatment of this model (both for one and $N$ fermion flavors) are in progress.

\vspace{1.5cm}

\noindent
{\Large
\bf
Acknowledgements}

\vspace{0.5cm}

J. T. L. was partially supported by Coordena\c c\~ao de Aperfei\c coamento de Pessoal de N\'{\i}vel Superior (CAPES/PICDT). L. A. M. was supported by Funda\c c\~ao de Amparo \`a Pesquisa do Estado de S\~ao Paulo (FAPESP). B. M. P. was supported in part by Conselho Nacional de Desenvolvimento Cient\'{\i}fico e Tecnol\'ogico (CNPq).

\pagebreak

\end{document}